\documentclass[aps,prd,preprint,showpacs,superscriptaddress]{revtex4-1}
\usepackage{amsmath}
\usepackage{latexsym}
\usepackage{amsfonts}
\usepackage{graphicx}
\usepackage{hyperref}

\begin{document}

\title{Modified Lyth bound and implications of BICEP2 results}

\author{Qing  Gao}
\email{gaoqing01good@163.com}
\affiliation{MOE Key Laboratory of Fundamental Quantities Measurement,
School of Physics, Huazhong University of Science and Technology,
Wuhan 430074,  P. R. China }

\author{Yungui Gong}
\email{yggong@mail.hust.edu.cn}
\affiliation{MOE Key Laboratory of Fundamental Quantities Measurement,
School of Physics, Huazhong University of Science and Technology,
Wuhan 430074,  P. R. China }

\author{Tianjun Li}
\email{tli@itp.ac.cn}
\affiliation{State Key Laboratory of Theoretical Physics
and Kavli Institute for Theoretical Physics China (KITPC),
Institute of Theoretical Physics, Chinese Academy of Sciences,
Beijing 100190, P. R. China}
\affiliation{School of Physical Electronics,
University of Electronic Science and Technology of China,
Chengdu 610054, P. R. China}

\begin{abstract}

To reconcile the BICEP2 measurement on the tensor-to-scalar ratio $r$ with Planck constraint,
a large negative running of scalar spectral index $n_s$ is needed. So the inflationary observable
such as $n_s$ should be expanded at least to the second-order slow-roll parameters for single-field
inflationary models. The large value of $r$ and the Lyth bound indicate that it is impossible to obtain
the sub-Planckian excursion for the inflaton. However, we derive an absolutely
minimal bound $\Delta\phi/M_{\rm Pl}>\sqrt{r/2}$
on the inflaton excursion for single-field inflationary models,
which can be applied to non-slow-roll inflationary models as well. This
bound excludes the possibility of the small-field inflation with $\Delta\phi<0.1 M_{\rm Pl}$
if the BICEP2 result on $r$ stands and it opens the window of
sub-Planckian excursion with $\Delta\phi<M_{\rm Pl}$ even
if $r$ is as large as $0.1$. To get the sub-Planckian excursion with $\Delta\phi<0.1 M_{\rm Pl}$,
our modified bound requires $r<0.02$.
Using a fifth-order polynomial potential as an explicit example, we show that it
not only agrees with the observational results, but also violates the Lyth bound.

\end{abstract}

\preprint{arXiv: 1405.6451}
\maketitle

\section{Introduction }

Inflation not only solves various problems in the standard big bang cosmology
such as the flatness, horizon and monopole problems, etc, but also provides
the seed of a large-scale structure
by the quantum fluctuation of the inflaton \cite{starobinskyfr, guth81, linde83, Albrecht:1982wi}.
Generic inflationary models predict that the spectrum of the density perturbation
is Gaussian, adiabatic, and almost scale invariant.  Besides the scalar perturbation,
the tensor perturbation is generated as well,
which gives the B-mode polarization
as a signature of the primordial gravitational wave.
The measurements of the cosmic microwave background radiation anisotropies provide a
strong test of inflationary models.
The Planck data of the temperature power spectrum \cite{planck13} in combination with
the nine years of WMAP polarization low-multipole likelihood data \cite{wmap9}
and the high-multipole spectra data from the Atacama Cosmology Telescope \cite{act13} and the South Pole Telescope \cite{spt11}
(Planck+WP+highL) constrained the scalar spectral index to be $n_s=0.960\pm 0.014$ and the tensor-to-scalar
ratio to be $r_{0.002} \le 0.11$ at the 95\% C.L. \cite{Ade:2013zuv,Ade:2013uln}.

However, the ground-based the BICEP2 experiment
has measured the tensor-to-scalar ratio to be $r=0.20^{+0.07}_{-0.05}$ at the 68\% C.L.
for the lensed-$\Lambda$CDM model in which the Universe contains a cosmological constant and cold dark matter, with
$r=0$ disfavoured at the $7.0\sigma$ level \cite{Ade:2014xna}
\footnote{Although the recent joint analysis \cite{Ade:2015tva} does not support this result,
the discussion and the main results of this paper are not affected.}.
The BICEP2 result on the tensor-to-scalar ratio is in tension with the Planck result.
If the running $n'_s=d\ln n_s/d\ln k$ of the spectral index is included, then the results between
the Planck and BICEP2 experiments can be consistent. With the running of spectral index,
the Planck+WP+highL data give
$n_s=0.9570\pm 0.0075$ and $n'_s=-0.022\pm 0.010$ at the 68\% C.L., and the tensor-to-scalar ratio is
constrained to be $r_{0.002}<0.26$ at the 95\% C.L.. The combined Planck+WP+highL+BICEP2
data give the constraints $n_s=0.9574^{+0.0073}_{-0.0074}$, $n_s'=-0.0292\pm 0.0096$ and $r_{0.002}=0.21^{+0.05}_{-0.06}$ at the 68\% C.L.

The large value of tensor-to-scalar ratio excludes a large class of inflationary models.
For example, the inflationary model with nonminimal coupling with gravity found that $n_s=1-2/N$
and $r=12/N^2$ \cite{Kallosh:2013tua}. If we take $N=50$,
then we get $r=0.0048$, so the model is excluded by the BICEP2 result.
The BICEP2 result also disfavors the small-field inflation
such as the hilltop inflation with the potential
$V(\phi)=V_0[1-(\phi/\mu)^p]$ \cite{Albrecht:1982wi,Boubekeur:2005zm},
the hybrid inflation~\cite{Linde:1993cn, Dvali:1994ms, Copeland:1994vg},
and many string-inspired models~\cite{Burgess:2013sla}.
On the other hand, the running of spectral index is at the order of at most $10^{-3}$ for the simple single-field inflation
because the scalar spectral index $n_s$ deviates
from $1$ at the order of $10^{-2}$. So the large negative running of
spectral index becomes a big challenge to single-field inflationary models \cite{Gong:2014cqa}.
From the naive analysis of the Lyth bound $\Delta\phi/M_{{\rm Pl}}>N\sqrt{r/8}$ \cite{Lyth:1996im}
with $M^2_{\rm Pl}=(8\pi G)^{-1}$, generically we need
large-field inflation if $r$ is in the order of $0.1$,
and then the validity of effective field theory becomes an issue
since higher-dimensional operators are not suppressed by the
reduced Planck scale.

Therefore, a successful single-field inflationary model should satisfy the following three criteria:
\begin{enumerate}

\item[C1.] The spectral index $n_s$ and the tensor-to-scalar ratio $r$ are consistent with the observational values.

\item[C2.]  A large negative running of spectral index $n'_s \sim -0.03$ is required if the running is included.

\item[C3.]  The Lyth bound should be violated so that the sub-Planckian excursion of the inflaton can be realized unless
the tensor-to-scalar ratio $r<0.001$.
\end{enumerate}

Inflationary models that satisfy the first condition C1 have been studied extensively
\cite{BenDayan:2009kv,Lizarraga:2014eaa,Harigaya:2014qza,Contaldi:2014zua,Collins:2014yua,Byrnes:2014xua,Anchordoqui:2014uua,Harigaya:2014sua,
Nakayama:2014koa,Zhao:2014rna,Cook:2014dga,Miranda:2014wga,Masina:2014yga,Hamada:2014iga,
Hertzberg:2014aha,Dent:2014rga,Joergensen:2014rya,Freese:2014nla,Czerny:2014wua,Ferrara:2014ima, Zhu:2014wda,
Okada:2014lxa, Ellis:2014rxa, Kawai:2014doa, Antusch:2014cpa,
Freivogel:2014hca, Bousso:2014jca, Kaloper:2014zba,Choudhury:2013iaa,*Choudhury:2014wsa,*Choudhury:2014kma,*Choudhury:2014uxa,Choi:2014aca,
Murayama:2014saa, McDonald:2014oza, Gao:2014fha, Li:2014owa, Li:2014xna, Ashoorioon:2014nta,*Ashoorioon:2013eia,*Ashoorioon:2009wa,*Ashoorioon:2011ki,Sloth:2014sga,
Kawai:2014doa,Kobayashi:2014rla,*Kobayashi:2014ooa,*Kobayashi:2014jga,Bastero-Gil:2014oga,DiBari:2014oja,Ho:2014xza,Hotchkiss:2011gz,
Dvali:2014ssa,Palti:2014kza,Ma:2014vua,Hamaguchi:2014mza,Creminelli:2014oaa,Kamionkowski:2014faa,
Carrillo-Gonzalez:2014tia,delCampo:2014toa,Marchesano:2014mla,Choi:2014dva,Oda:2014rpa,Yonekura:2014oja,Higaki:2014sja,Bamba:2014jia,Hebecker:2014eua}.
Natural inflation models with sinusoidal potential can easily satisfy the first two conditions C1 and C2 \cite{Czerny:2014wua,Higaki:2014sja}.
With the help of two small decay constants, an effective large decay constant is realized for the natural inflation in string theory
so that the condition C3 may be avoided \cite{Long:2014dta,Ben-Dayan:2014zsa}.
For the inflationary model building, it seems that supergravity
theory is a natural framework \cite{Freedman:1976xh,*Deser:1976eh,Antusch:2009ty,*Antusch:2011ei}.
However, supersymmetry-breaking scalar masses in a generic supergravity theory
are of the same order as the gravitino mass, inducing
the so-called $\eta$ problem~\cite{Copeland:1994vg,Stewart:1994ts,*adlinde90,*Antusch:2008pn,*Yamaguchi:2011kg,*Martin:2014vha,Lyth:1998xn},
where all the scalar masses are of the order of the Hubble parameter
due to the large vacuum energy density during inflation~\cite{Goncharov:1984qm}.
There are two elegant solutions: no-scale supergravity~\cite{Cremmer:1983bf,*Ellis:1983sf,*Ellis:1983ei,*Ellis:1984bm,*Lahanas:1986uc,
Ellis:1984bf, Enqvist:1985yc, Ellis:2013xoa, Ellis:2013nxa, Li:2013moa, Ellis:2013nka}
and shift symmetry in the K\"ahler potential~\cite{Kawasaki:2000yn, Yamaguchi:2000vm,
Yamaguchi:2001pw, Kawasaki:2001as, Kallosh:2010ug, Kallosh:2010xz, Nakayama:2013jka,
Nakayama:2013txa, Takahashi:2013cxa, Li:2013nfa}.

In this paper, we derive the modified Lyth bound $\Delta\phi/M_{\rm Pl}>\sqrt{r/2}$
on the inflaton excursion for a single-field inflation,
which is applicable to any inflationary model. This
bound excludes the possibility of small-field inflation with $\Delta\phi<0.1 M_{\rm Pl}$
if the BICEP2 result on $r$ stands, and it opens the window for the sub-Planckian excursion with $\Delta\phi<M_{\rm Pl}$ even
if $r$ is as large as $0.1$. The modified Lyth bound also tells us that the sub-Planckian
excursion with $\Delta\phi<0.1 M_{\rm Pl}$ requires $r<0.002$.
So the constraint on the field excursion with the condition C3 will be replaced by the modified Lyth bound.
Then, we explicitly construct a polynomial inflationary model that satisfies
all the conditions C1, C2, and C3.
The model not only agrees with the observational results
but also supports the modified Lyth bound.

\section{Polynomial Potential and Supergravity Model Building}

For a given K\"ahler potential $K$ and a superpotential $W$
in the supergravity theory, we have the scalar potential
\begin{equation}
\label{sgp}
V=e^K\left((K^{-1})^{i}_{\bar{j}}D_i W D^{\bar{j}} \overline{W}-3|W|^2 \right)~,~
\end{equation}
where $(K^{-1})^{i}_{\bar{j}}$ is the inverse of the K\"ahler metric
$K_{i}^{\bar{j}}=\partial^2 K/\partial \Phi^i\partial{\bar{\Phi}}_{\bar{j}}$, and $D_iW=W_i+K_iW$.
The kinetic term for the scalar field $\Phi^i$ is
\begin{equation}
{\cal L} ~=~ K_{i}^{\bar{j}} \partial_{\mu} \Phi^i \partial^{\mu} {\bar \Phi}_{\bar{j}}~.~\,
\end{equation}
With the help of two superfields $\Phi$ and $X$, we consider the K\"ahler potential and superpotential
\begin{gather}
\label{KP-A}
K=\frac{1}{2}(\Phi+{\bar \Phi})^2+X{\bar X}-\delta(X{\bar X})^2,\\
\label{SP-A}
W=Xf(\Phi),
\end{gather}
so that the K\"ahler potential $K$ is invariant under the shift
symmetry~\cite{Kawasaki:2000yn, Yamaguchi:2000vm,
Yamaguchi:2001pw, Kawasaki:2001as, Kallosh:2010ug, Kallosh:2010xz, Nakayama:2013jka,
Nakayama:2013txa, Takahashi:2013cxa, Li:2013nfa}
\begin{eqnarray}
\Phi\rightarrow\Phi+iCM_{\rm Pl}~,~\,
\label{SSymmetry-A}
\end{eqnarray}
with $C$ a dimensionless real parameter. Because
the K\"ahler potential $K$ is a function of $\Phi+\Phi^{\dagger}$, it is independent of
the imaginary part of $\Phi$. Substituting the K\"ahler potential $K$ (\ref{KP-A})
and superpotential $W$ (\ref{SP-A}) into the scalar potential (\ref{sgp}),
the scalar potential becomes
\begin{equation}
V= e^K\left[ \left|(\Phi + {\bar \Phi})Xf(\Phi)+X \frac{\partial f(\Phi)}{\partial \Phi}\right|^2-3|Xf(\Phi)|^2
+|({\bar X}-2\delta X{\bar X}^2)Xf(\Phi)+f(\Phi)|^2\right].
\end{equation}
Because there is no imaginary component ${\rm Im} [\Phi]$ of $\Phi$ in the K\"ahler potential
due to the shift symmetry, the potential along ${\rm Im} [\Phi]$ is very flat and then
${\rm Im} [\Phi]$ is a natural inflaton candidate.
As we know from the previous studies~\cite{Kallosh:2010ug, Kallosh:2010xz, Li:2013nfa},
 the real component ${\rm Re} [\Phi]$ of $\Phi$ and $X$ can be
stabilized at the origin during inflation, {\it i.e.}, ${\rm Re} [\Phi]=0$ and $X=0$.
Therefore, with ${\rm Im} [\Phi]=\phi/{\sqrt 2}$, we obtain the inflaton potential
\begin{eqnarray}
V~=~|f(\phi/{\sqrt 2})|^2.
\end{eqnarray}
Considering a polynomial function $f(\Phi)~=~\sum_{n=0} (-i{\sqrt 2})^n a'_n \Phi^n$,
we obtain the polynomial inflaton potential
\begin{equation}
V(\phi)=\left|\sum_{n=0}  a'_n (\phi/M_{\rm Pl})^n\right|^2.
\end{equation}
To simplify the above potential, we define
\begin{eqnarray}
V_0~\equiv~ |a'_0|^2~,~~a_n~\equiv~\frac{a'_n}{a'_0}~.~\,
\end{eqnarray}
The inflaton potential can be rewritten as follows
\begin{equation}
\label{SUSY-PolyP}
V(\phi)=V_0 \left|a_0+\sum_{n=1}  a_n (\phi/M_{\rm Pl})^n\right|^2.
\end{equation}
In particular, we want to emphasize $a_0=1$. Let us consider the polynomial potential of the inflaton
\begin{equation}
V(\phi) = V_0\left[1+\sum_{m=1} \lambda_m \left(\frac{\phi-\phi_*}{M_{\rm Pl}}\right)^m \right]~,~\,
\end{equation}
where the subscript * means the value at the horizon crossing.
Without loss of generality, we will take $\phi_*=0$.

If the above polynomial potential is from the scalar potential in Eq.~(\ref{SUSY-PolyP})
for the supergravity model building, we have
\begin{equation}
\lambda_m ~=~\sum_{i< m/2} 2 a_i a_{m-i}
\end{equation}
for odd $m$ and
\begin{equation}
\lambda_m ~=~\sum_{i< m/2} 2 a_i a_{m-i} +  a_{m/2}^2
\end{equation}
for even $m$.

\section{Lyth bound in slow-roll inflation}

In terms of the slow-roll parameters
\begin{gather}
\label{slow1}
\epsilon(\phi)=\frac{M_{\rm Pl}^2V_\phi^2}{2V^2},\\
\label{slow2}
\eta(\phi)=\frac{M_{\rm Pl}^2V_{\phi\phi}}{V},\\
\label{slow3}
\xi^2(\phi)=\frac{M_{\rm Pl}^4V_\phi V_{\phi\phi\phi}}{V^2},
\end{gather}
the scalar spectral index and its running are
given by \cite{Lyth:1998xn,Stewart:1993bc}
\begin{gather}
\label{nsdef}
n_s\approx 1+2\eta-6\epsilon+2\left[\frac{1}{3}\eta^2+(8C-1)\epsilon \eta-\left(\frac{5}{3}+12C\right)\epsilon^2-\left(C-\frac{1}{3}\right)\xi^2\right],\\
\label{rundef}
n_s'=16\epsilon\eta-24\epsilon^2-2\xi^2,
\end{gather}
where $V_\phi=dV(\phi)/d\phi$, $V_{\phi\phi}=d^2V(\phi)/d\phi^2$,
$V_{\phi\phi\phi}=d^3V(\phi)/d\phi^3$, and $C=-2+\ln 2+\gamma \simeq -0.73$ with $\gamma$ the Euler--Mascheroni constant.
The tensor spectral index and tensor-to-scalar ratio are \cite{Lyth:1998xn,Stewart:1993bc}
\begin{gather}
\label{ntdef}
n_t=-2\epsilon\left[1+\left(4C+\frac{11}{3}\right)\epsilon-2\left(\frac{2}{3}+C\right)\eta\right]
\approx -2\epsilon,\\
\label{rdef}
r=16 \epsilon \left[ 1+2\left(C-\frac{1}{3}\right)(2\epsilon-\eta) \right]\approx 16\epsilon.
\end{gather}
Note that the quantities $\epsilon$, $\eta$ and $\xi^2$ in Eqs. (\ref{nsdef})-(\ref{rdef}) are evaluated
at the horizon crossing $k=aH$ or $\phi_*$ for a given scale $k$.
As discussed in Ref. \cite{Gao:2014yra}, the observational result of large running
requires us to consider the second-order correction
to the scalar spectral index $n_s$ in Eq. (\ref{nsdef})
because $\xi^2$ has the same order as $\epsilon$ and $\eta$ and the main contribution
to the running of spectral index comes from $\xi^2$.
The number of $e$-folds before the end of inflation is given by
\begin{equation}
\label{efolddef}
N(\phi)=\int_t^{t_e}Hdt\approx \frac{1}{M_{\rm Pl}}\int_{\phi_e}^\phi\frac{d\phi}{\sqrt{2\epsilon(\phi)}},
\end{equation}
where the value $\phi_e$ of inflaton at the end of inflation is defined by
$\epsilon(\phi_e)=1$ or $\eta(\phi_e)=1$.
If $\epsilon(\phi)$ is a monotonic function of $\phi$ during inflation,
 we have $\epsilon(\phi)>\epsilon(\phi_*)=\epsilon=r/16$, and then get the Lyth bound \cite{Lyth:1996im}
\begin{equation}
\Delta \phi \equiv |\phi_*-\phi_e| > \sqrt{2\epsilon}\, N(\phi_*) M_{\rm Pl}=\sqrt{r/8}\, N(\phi_*) M_{\rm Pl}.
\end{equation}
Note that the Lyth bound is not related to the spectral index $n_s$ and it
holds for general slow-roll inflationary models whether they are consistent
with observational constraints or not.
Therefore, $r=0.21$ requires large-field inflation due to $\Delta\phi > 9.72 M_{\rm Pl}$
for $N=60$. If we take $r=0.1$, then the Lyth bound requires
$\Delta\phi>6.71 M_{\rm Pl}$ for $N=60$.
To get the sub-Planckian excursion for the inflaton, the Lyth bound requires
 $r\lesssim 0.001$. Therefore, to reconcile the BICEP2 result and small-field inflation,
the Lyth bound must be violated. In short,
$\epsilon(\phi)$ should not be a monotonic function and needs to
have at least one minimum between $\phi_*$ and $\phi_e$ \cite{BenDayan:2009kv,Hebecker:2013zda},
the inflaton value of which is defined as $\phi_{\rm min}$.
It was argued in Ref. \cite{Antusch:2014cpa} that it is impossible to achieve
$\Delta\phi<M_{\rm Pl}$ for single-field inflationary models because $\Delta\phi/M_{\rm Pl}\gtrsim \sqrt{r/8}/\langle\eta-2\epsilon\rangle$,
where $\langle\eta-2\epsilon\rangle$ is the mean of $\eta-2\epsilon$ between $\phi_{\rm min}$ and $\phi_*$.
In Ref.~\cite{BenDayan:2009kv}, a large tensor-to-scalar ratio and large negative running were obtained by
a single polynomial potential. In particular, they found that $n_s=0.96$, $r=0.1$, $n_s'=-0.07$, and
$\Delta\phi(N=60)\sim M_{\rm Pl}$.  Furthermore, it was argued
that $\Delta\phi$ lies in a narrow range below $M_{\rm Pl}$
in Refs. \cite{Choudhury:2014kma,German:2014qza}. In Ref. \cite{Efstathiou:2005tq}, it was derived
that $\Delta\phi/M_{\rm Pl}\approx 6r^{1/4}$. It is evident that the Lyth bound
is not the lowest bound and a weaker bound exists; the main goal of this work is to
derive the lowest bound, and present some concrete examples.

To violate the Lyth bound and get the sub-Planckian excursion, the main contribution
to $N$ must come from $1/\sqrt{2\epsilon(\phi_{\rm min})}$,
so $\epsilon(\phi_{\rm min})<1/(2N)^2\sim 10^{-4}\ll\epsilon(\phi_*)$
and $\Delta\epsilon(\phi)=\epsilon(\phi_*)-\epsilon(\phi_{\rm min})\sim \epsilon$.
Note that
\begin{gather}
\label{depsilon1}
M_{\rm Pl}\left|\frac{d\epsilon(\phi)}{d\phi}\right|=\sqrt{2\epsilon(\phi)}|\eta(\phi)-2\epsilon(\phi)|,\\
\label{depsilon2}
M^2_{\rm Pl}\frac{d^2\epsilon(\phi)}{d\phi^2}=\eta^2(\phi)-10\epsilon(\phi)\eta(\phi)+12\epsilon^2(\phi)+\xi^2(\phi).
\end{gather}
So for slow-roll inflation,  before the scalar field reaches $\phi_{\rm min}$,
both $\epsilon_\phi=d\epsilon(\phi)/d\phi$ and $\epsilon_{\phi\phi}=d^2\epsilon(\phi)/d\phi^2$ are small.
If the excursion of scalar field $\Delta\phi/M_{\rm Pl}<\sqrt{2\epsilon}$,
then to get $\Delta\epsilon(\phi)\sim \epsilon$, the higher-order derivatives must be large
and give the major contribution to $\Delta\epsilon(\phi)$. When the higher-order derivatives are large,
the slow-roll condition may not be satisfied, and the
higher-order corrections to the scalar spectral index cannot be neglected. For example,
the third-order correction to $n_s$ is \cite{Gong:2001he,Zhu:2014aea}
\begin{equation}
\label{ns3rd}
\begin{split}
&\left(-96C^2-\frac{104}{3}C-\frac{3734}{9}+44\pi^2\right)\epsilon^3+\left(96C^2-\frac{4}{3}C+\frac{1190}{3}-44\pi^2\right)\epsilon^2\eta\\
&+\left(-16C^2+12C-\frac{742}{9}+\frac{28\pi^2}{3}\right)\epsilon\eta^2+\frac{4}{9}\eta^3+\left(-12C^2+4C-\frac{98}{3}+4\pi^2\right)\epsilon\xi^2\\
&+\left(C^2-\frac{8}{3}C+\frac{28}{3}-\frac{13\pi^2}{12}\right)\eta\xi^2+\left(C^2-\frac{2}{3}C+\frac{2}{9}-\frac{\pi^2}{12}\right)\sigma^3,
\end{split}
\end{equation}
where $\sigma^3=M^6_{\rm Pl}(V_\phi)^2V_{\phi\phi\phi\phi}/V^3$.
The last term may contribute to $n_s$ if $\epsilon(\phi)$ changes fast. Therefore, it is impossible to get
sub-Planckian excursion with $\Delta\phi<0.1 M_{\rm Pl}$
for single-field slow-roll inflation if $r$ is as large as $0.1$.
Integrating Eq. (\ref{depsilon1}), we get
\begin{equation}
\sqrt{2\epsilon(\phi_*)}-\sqrt{2\epsilon(\phi_{\rm min})}=\int_{\phi_*}^{\phi_{\rm min}}d\phi\, [\eta(\phi)-2\epsilon(\phi)]/M_{\rm Pl}.
\end{equation}
Since $N=50-60$, $\epsilon(\phi_{\rm min})\ll \epsilon(\phi_*)$, so we obtain \cite{Antusch:2014cpa}
\begin{equation}
\label{abound}
\frac{\Delta\phi}{M_{\rm Pl}}\gtrsim \frac{\sqrt{2\epsilon}}{\langle\eta-2\epsilon\rangle}
=\frac{\sqrt{2\epsilon}}{\sqrt{2\epsilon}-\sqrt{2\epsilon(\phi_{\rm min})}}\frac{\phi_{\rm min}-\phi_*}{M_{\rm Pl}}
\approx \frac{\phi_{\rm min}-\phi_*}{M_{\rm Pl}}.
\end{equation}
The above bound is trivial since it is just the difference between $\phi_{\rm min}$ and $\phi_*$,
and it is true only when a minimum for $\epsilon(\phi)$ exists in the region between $\phi_*$ and $\phi_e$. The bound
is also useless because it tells us nothing even if we have the observational information on $n_s$ and $r$.
To obtain the concrete value of $\langle\eta-2\epsilon\rangle$, we need to calculate both $\phi_{\rm min}$ and
$\phi_*$, so it is not easy to calculate the average value of $\eta-2\epsilon$ and get
the value of the bound (\ref{abound}) in general. In particular, for this bound,
we still have no idea whether the sub-Planckian excursion can be
realized when observational constraints on $n_s$ and $r$ are known.
To get enough $N$, $\phi_{\rm min}$ is usually close to $\phi_e$, and the potential changes rapidly after $\phi_{\rm min}$.
The slow-roll parameters $\epsilon$ and $\eta$ are smaller than 1, and $\phi_{\rm min}$ is not far away from $\phi_e$, so
we expect that $\langle\eta-2\epsilon\rangle<1/2$.
Therefore, we propose the absolutely minimal modified Lyth bound on $\Delta\phi$,
\begin{equation}
\label{gglbound}
\frac{\Delta\phi}{M_{\rm Pl}}>\sqrt{8\epsilon}=\sqrt{\frac{r}{2}}~.~\,
\end{equation}
Apparently the bound (\ref{gglbound}) is smaller than the bound (\ref{abound}),
and it is more practical and useful since
it depends only on the slow-roll parameter $\epsilon$, or the observable $r$.
The bound holds even if $\epsilon(\phi)$ has no minimum between $\phi_*$ and $\phi_e$.
The modified Lyth bound is lower,
universal, and model independent in the sense that it involves the slow-roll parameter $\epsilon$ only.
Although the number of $e$-folds $N(\phi_*)$ before the end of inflation is absent in the above
bounds (\ref{abound}) and (\ref{gglbound}), the results are not independent of $N(\phi_*)$
because they hold under the assumption that $\epsilon(\phi_{\rm min})\ll \epsilon(\phi_*)$,
which is based on large $N$.
The modified Lyth bound (\ref{gglbound}) holds for any single-field inflationary model and
is independent of the conditions C1 and C2.
With the modified Lyth bound, it is possible to get the sub-Planckian excursion for the inflaton
with $\Delta\phi<M_{\rm Pl}$
even though $r$ is as large as $0.1$.
We want to emphasize that it is very difficult to saturate this bound.
The modified Lyth bound (\ref{gglbound}) tells us that $r$ should satisfy $r<0.02$ to get $\Delta\phi<0.1 M_{\rm Pl}$.

For a fifth polynomial potential with $\lambda_1=-0.1581$, $\lambda_2=0.003433$,
$\lambda_3=-0.01054$, $\lambda_4=0.9115$ and $\lambda_5=-0.97$, we get
$\Delta\phi=\phi_e=0.67 M_{\rm Pl}$ with $N=57.5$ and $r=0.2$.
For this potential, $\langle\eta-2\epsilon\rangle=0.28$, and
the bound (\ref{abound}) requires $\Delta\phi>0.56 M_{\rm Pl}$, the Lyth bound requires $\Delta\phi>9.1 M_{\rm Pl}$,
and our modified bound requires $\Delta\phi>0.32 M_{\rm Pl}$, which is lower than the bound (\ref{abound}).
The result $\Delta\phi=0.67 M_{\rm Pl}$
satisfies both the bounds (\ref{abound}) and (\ref{gglbound}) and violates the Lyth bound. If we take
$\lambda_1=-0.1581$, $\lambda_2=0.003433$,
$\lambda_3=-0.01054$, $\lambda_4=5.0$, and $\lambda_5=-9.4728$, we get $\Delta\phi=0.36 M_{\rm Pl}$ with $N=57.1$ and $r=0.2$.
For this case, $\langle\eta-2\epsilon\rangle=0.4994$, both bounds (\ref{abound}) and (\ref{gglbound}) require
$\Delta\phi>0.32 M_{\rm Pl}$, and the Lyth bound requires $\Delta\phi>9.03 M_{\rm Pl}$. The model also
satisfies both the bounds (\ref{abound}) and (\ref{gglbound}) and violates the Lyth bound.
If we take $r=0.01$, $\lambda_1=-0.0353553$, $\lambda_2=-0.00802833$,
$\lambda_3=-0.00235702$, $\lambda_4=57.0$, and $\lambda_5=-396.072$, we get $\Delta\phi=\phi_e=0.099M_{\rm Pl}$ and $N=58.5$.
The corresponding Lyth bound is $\Delta\phi>2.1 M_{\rm Pl}$, the modified Lyth bound (\ref{gglbound}) is
$\Delta\phi>0.07 M_{\rm Pl}$, and the bound (\ref{abound}) is $\Delta\phi>0.087 M_{\rm Pl}$ with $\langle\eta-2\epsilon\rangle=0.41$.
The above result is derived based on the slow-roll approximation, and in general, we expect the inflaton
excursion is larger than the modified Lyth bound (\ref{gglbound}) if we solve the exact dynamical
equation either analytically or numerically, especially for the nonstandard slow-roll potentials.
For the fifth polynomial potential with $\lambda_1=-0.1581$, $\lambda_2=0.003433$,
$\lambda_3=-0.01054$, $\lambda_4=0.9115$, and $\lambda_5=-0.97$, the numerical result gives $\Delta\phi=\phi_e=1.14M_{\rm Pl}$
and $N=59.2$. For the potential with $\lambda_1=-0.1581$, $\lambda_2=0.003433$,
$\lambda_3=-0.01054$, $\lambda_4=5.0$, and $\lambda_5=-9.4728$, the numerical result gives $\Delta\phi=\phi_e=0.69M_{\rm Pl}$
and $N=59.2$. For the potential with $\lambda_1=-0.0353553$, $\lambda_2=-0.00802833$,
$\lambda_3=-0.00235702$, $\lambda_4=57.0$, and $\lambda_5=-396.072$, the numerical result gives $\Delta\phi=\phi_e=0.29M_{\rm Pl}$
and $N=59.8$.

\section{Polynomial Potential}

Now, let us show how to construct an inflationary model that satisfies all three conditions C1-C3
with the polynomial potential as an example.
For simplicity, we assume $M_{\rm Pl}=1$ and
denote the magnitudes of the inflaton $\phi$ at the
horizon crossing and the end of inflation as $\phi_*$ and $\phi_e$, respectively.
For the polynomial potential, the slow-roll parameters at the horizon crossing $\phi_*$ are
\begin{equation}
\label{polyslow1}
\epsilon=\frac{r}{16}=\frac{\lambda_1^2}{2},\ \eta=2\lambda_2,\
\xi^2=6\lambda_1\lambda_3.
\end{equation}
From the observational constraint on $r$, we can get the coefficient $\lambda_1$.
As we discussed above, the main contribution to the running
of the scalar spectral index comes from $\xi^2$. So the observational constraints on $\epsilon$ and $n_s'$
give the coefficient $\lambda_3$ \cite{BenDayan:2009kv},
\begin{equation}
\label{polypars1}
\lambda_1=-\sqrt{2\epsilon}=-\sqrt{\frac{r}{8}},\
\lambda_3\approx \frac{n_s'}{3\sqrt{2r}}.
\end{equation}
Once $\epsilon$ and $\xi^2$ are known, the slow-roll parameter $\eta$ is determined from the scalar spectral index (\ref{nsdef}),
and the coefficient $\lambda_2$ is
\begin{equation}
\label{polypars2}
\lambda_2\approx \frac{n_s-1}{4}+\frac{3r}{32}-\frac{1}{4}(C-1/3)n_s'.
\end{equation}

For $m=3$, if we take $\lambda_1=-0.162$, $\lambda_2=0.0016$, and $\lambda_3=-0.0132$,
we get $n_s=0.957$, $r=0.21$, $n_s'=-0.0292$, and $\phi_e=2.7M_{\rm Pl}$.
Because $\epsilon$ is a monotonic function, the number of $e$-folds before
the end of inflation is $N=9.12$, which is not large enough
to solve the horizon problem. Therefore, we need to introduce a few more terms $\lambda_m$ with $m > 3$ so that
we have a high enough number of $e$-folds. If we add one more
term and consider $m=4$, we find that the slow-roll parameters
are always smaller than $1$ if $\lambda_4$ is too small and that $\phi_{\rm min}$ decreases as
$\lambda_4$ increases.
Because the third-order slow-roll parameter $\sigma^3$ is proportional to $\lambda_4$,
the slow-roll condition requires that $\lambda_4$ be small. Therefore,
$\lambda_4$ lies in a narrow region. However, for those values of $\lambda_4$, $\epsilon(\phi_{\rm min})=0$
and near $\phi_{\rm min}$,
$\epsilon(\phi)\approx \epsilon_{\phi\phi}(\phi_{\rm min})(\phi-\phi_{\rm min})^2/2$,
the integral $\int 1/\sqrt{2\epsilon(\phi)}d\phi$ is logarithm divergent, and then we need to consider more terms.
From the number of freedom counting,
we might only need to introduce at most two terms, for example,
the $\lambda_4$ and $\lambda_5$ terms. The coefficients $\lambda_4$ and $\lambda_5$ are then determined from
$N(\phi=\phi_*)=60$ and $\epsilon(\phi_e)=1$ (or $\eta(\phi_e)=1$) \cite{BenDayan:2009kv}. Additionally,
we require that the potential has a flat plateau so that $\epsilon(\phi)$ has a minimum and
$\epsilon(\phi_{\rm min})$ is close to 0. At $\phi_{\rm min}$, $\eta\approx 0$, so we may express
$\lambda_5$ in terms of $\phi_{\rm min}$ and $\lambda_4$, and then $\epsilon(\phi_{\rm min})$ is a function of $\lambda_4$.
The number of $e$-folds before
the end of inflation is usually between 50 and 60. To get enough $e$-folds, we require $\epsilon(\phi_{\rm min})$ to be
around $10^{-6}$, and then $\lambda_4$ is determined once $\phi_{\rm min}$ is given.

Following the procedure discussed above, we
construct inflationary models that are consistent with the observational constraints.
In particular, we consider the inflaton potential with
 $\lambda_1=-0.162$, $\lambda_2=0.00161$, $\lambda_3=-0.0132$, $\lambda_4=0.01$,
and $\lambda_5=-0.00146$. By using the slow-roll formula, we get $n_s=0.957$ with the second-order correction (\ref{nsdef}),
$r=0.21$, $n_s'=-0.0292$, $N=56.4$, and $\phi_e=4.42$. So $\Delta\phi=4.42<N\sqrt{2\epsilon}=9.13$.
For the potential with $\lambda_1=-0.0354$, $\lambda_2=-0.0078$, $\lambda_3=-0.00024$, $\lambda_4=1.398$,
and $\lambda_5=-2.766$, we get $n_s=0.9655$, $r=0.01$, $n_s'=-0.0003$, $N=58.7$, and $\phi_e=0.41$, which
is much less than the Lyth bound $\Delta\phi>2.08$.

\begin{figure}[htp]
\centerline{\includegraphics[width=0.4\textwidth]{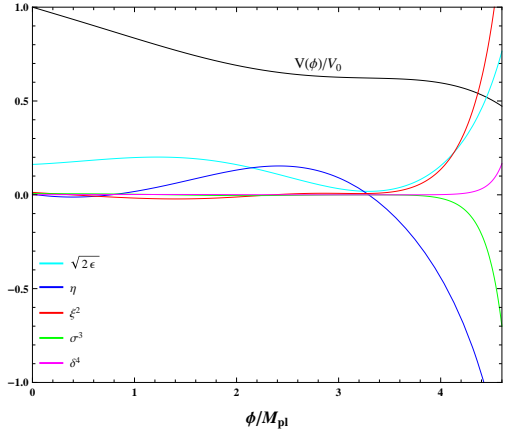}}
\caption{The behavior of the potential $V(\phi)$ and the corresponding
slow-roll parameters $\epsilon(\phi)$, $\eta(\phi)$, $\xi^2(\phi)$, $\sigma^3(\phi)$,
and $\delta^4(\phi)=M^8_{\rm Pl}(V_\phi)^3V_{\phi\phi\phi\phi\phi}/V^4(\phi)$ for the polynomial potential
with the coefficients $\lambda_1=-0.162$, $\lambda_2=0.00161$, $\lambda_3=-0.0132$, $\lambda_4=0.01$,
and $\lambda_5=-0.00146$.}
\label{slowroll}
\end{figure}

\begin{figure}[htp]
\centerline{\includegraphics[width=0.4\textwidth]{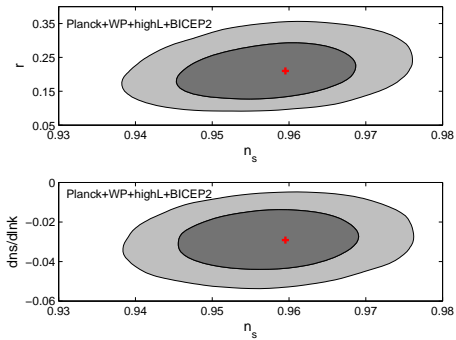}}
\caption{The ``+" stands for the results on $n_s$, $r$, and $n_s'$ from the polynomial potential
with $\lambda_1=-0.162$, $\lambda_2=0.00161$, $\lambda_3=-0.0132$, $\lambda_4=0.01$,
and $\lambda_5=-0.00146$. We also show the 68\%
 and 95\% contours constrained from the combination of Planck+WP+highL+BICEP2 data.}
\label{plkb2}
\end{figure}

To understand why the polynomial potential we constructed violates the Lyth bound but is
consistent with the observational results, we plot the potential and
the slow-roll parameters in Fig. \ref{slowroll}. We also show the slow-roll results and the
observational contours in Fig. \ref{plkb2}.
At the horizon crossing $\phi_*$, the potential has a large slope, so the slow-roll parameter
$\epsilon(\phi)$ is relatively large at $\phi_*$, and the derived $n_s$, $n_s'$, and $r$
are consistent with the observations. After the horizon crossing, the potential becomes very flat,
and $\epsilon(\phi)$ decreases to be very small. Near the end of inflation, the potential changes quickly,
and the slow-roll parameter $\epsilon(\phi)$ or $\eta(\phi)$ quickly increases to 1. Therefore, in principle,
all three conditions C1--C3 can
be satisfied if the potential has the above property.

Here, we construct the potential by using
the slow-roll conditions. If the slow-roll conditions are not satisfied, we need to solve
the Mukhanov--Sasaki equation numerically \cite{Mukhanov:1985rz,Sasaki:1986hm},
\begin{equation}
\label{mseq1}
\frac{d^2 v_k}{d\tau^2}+\left(k^2-\frac{1}{z}\frac{d^2z}{d\tau^2}\right)v_k=0,
\end{equation}
and calculate the power spectrum
\begin{equation}
\label{pwrspectrum1}
{\mathcal P}_{\mathcal R}=\frac{k^3}{2\pi^2}\left|\frac{v_k}{z}\right|^2=A_S\left(\frac{k}{k_*}\right)^{n_s-1+\frac{1}{2}n_s'\ln(k/k_*)+\cdots},
\end{equation}
where $z=a{\mathcal H}^{-1}d\phi/d\tau$, ${\mathcal H}=a^{-1}da/d\tau$, and the conformal time $d\tau=dt/a$.
By doing so, Ben--Dayan and Brustein obtained the
sub-Planckian excursion with $\Delta\phi\sim 0.5M_{\rm Pl}$ \cite{BenDayan:2009kv}.
For the potential we considered above, the numerical solution to the Mukhanov--Sasaki equation gives
$n_s=0.9587$, $n_s'=-0.0297$, and $r=0.2$ for $k_*=0.002$ Mpc$^{-1}$, which are consistent with the slow-roll results.
For this model, we also get $\Delta\phi=5.07~M_{\rm Pl}$ and $N=59.1$.  As expected, the numerical
results on $\Delta\phi$ and $N$ are larger than the slow-roll results.

\section{Discussions and Conclusion}

Whatever the conditions C1 and C2 are satisfied or not, both the Lyth bound and the modified Lyth bound hold for simple
slow-roll inflationary models with $N=50-60$. If $r$ is at the order of $0.1$, then the Lyth bound
tells us that $\Delta\phi\gtrsim 5.6M_{\rm Pl}$, and the sub-Planckian excursion for the inflaton is impossible.
To get the sub-Planckian excursion, the Lyth bound requires that $r\lesssim 10^{-3}$. On the other hand,
the Lyth bound is based on the assumption that $\epsilon(\phi)$ is a monotonic function, so it can be
violated if $\epsilon(\phi)$ is not a monotonic function. We have derived the absolutely minimal modified Lyth bound
$\Delta\phi/M_{\rm Pl}>\sqrt{8\epsilon}=\sqrt{r/2}$ by requiring that $N(\phi_*)$ be large. With the modified Lyth bound,
if $r=0.2$, then it is possible to get the sub-Planckian excursion for the inflaton since $\Delta\phi>0.32M_{\rm Pl}$.
For example, if we choose $\lambda_1=-0.1581$, $\lambda_2=0.00343$, $\lambda_3=-0.010541$, $\lambda_4=5.0$,
and $\lambda_5=-9.47$, we get $\Delta\phi=0.36 M_{\rm Pl}$ with $N=57.1$
while the Lyth bound requires that the model satisfies $\Delta\phi>9.03 M_{\rm Pl}$.
So the modified Lyth bound opens the possibility that all three conditions C1, C2, and C3 can be satisfied.
We provided a procedure to construct models that satisfy the conditions C1 and C2
by using the polynomial potential as an example. The model we considered in this work
gives $n_s=0.957$, $n_s'=-0.0292$, $r=0.21$, $\Delta\phi=4.42M_{\rm Pl}$ and $N=56.4$
by using the slow-roll formulas, and the more accurate numerical calculation gives
$n_s=0.9587$, $n_s'=-0.0297$, $r=0.2$, $N=59.1$, and $\Delta\phi=5.07M_{\rm Pl}$ for $k_*=0.002$ Mpc$^{-1}$.  So it
satisfies the conditions C1--C3. Furthermore, the model supports the modified Lyth bound.
The modified Lyth bound (\ref{gglbound}) is further supported by the work in Ref. \cite{BenDayan:2009kv} with $\Delta\phi\sim 0.5M_{\rm Pl}$.
Even if $r$ is found to be much smaller and the inflaton is a small field,
 the modified Lyth bound still limits the inflaton excursion.
For example, to get $\Delta\phi<0.1M_{\rm Pl}$, the modified Lyth bound (\ref{gglbound}) only requires $r<0.002$.
By using the slow-roll formulas, we constructed a polynomial potential with
$\lambda_1=-0.0354$, $\lambda_2=-0.0078$, $\lambda_3=-0.00024$, $\lambda_4=1.398$,
and $\lambda_5=-2.766$ to give $n_s=0.9655$, $r=0.01$, $n_s'=-0.0003$, and $N=58.7$. The corresponding
inflaton excursion
is $\Delta\phi=0.41 M_{\rm Pl}$.

\begin{acknowledgments}

This research was supported in part by the Natural Science
Foundation of China under Grants No. 10821504, No. 11075194, No. 11135003, No. 11175270, No. 11275246, No. 11475065, and No. 11475238;
the National Basic Research Program of China (973 Program) under Grant No. 2010CB833000 (T.L.);
the Program for New Century Excellent Talents in University under Grant No. NCET-12-0205;
and the Fundamental Research Funds for the Central Universities under Grant No. 2013YQ055.

\end{acknowledgments}


\begin{thebibliography}{100}

\bibitem{starobinskyfr}
A.~A. Starobinsky,
 Phys. Lett. B. {\bf 91}, 99 (1980).

\bibitem{guth81}
A.~H. Guth,
 Phys. Rev. D {\bf 23}, 347 (1981).

\bibitem{linde83}
A.~D. Linde,
 Phys. Lett. B {\bf 129}, 177 (1983).

\bibitem{Albrecht:1982wi}
A.~Albrecht and P.~J. Steinhardt,
 Phys. Rev. Lett. {\bf 48}, 1220 (1982).

\bibitem{planck13}
P.~Ade {\em et~al.} (Planck Collaboration),
 Astron. Astrophys. {\bf 571}, A1 (2014).

\bibitem{wmap9}
G.~Hinshaw {\em et~al.} (WMAP Collaboration),
 Astrophys. J. Suppl. {\bf 208}, 19 (2013).

\bibitem{act13}
S.~Das {\em et~al.},
 JCAP {\bf 1404}, 014 (2014).

\bibitem{spt11}
R.~Keisler {\em et~al.},
 Astrophys. J. {\bf 743}, 28 (2011).

\bibitem{Ade:2013zuv}
P.~Ade {\em et~al.} (Planck Collaboration),
 Astron. Astrophys. {\bf 571}, A16 (2014).

\bibitem{Ade:2013uln}
P.~Ade {\em et~al.} (Planck Collaboration),
 Astron.Astrophys. {\bf 571}, A22 (2014).

\bibitem{Ade:2014xna}
P.~Ade {\em et~al.} (BICEP2 Collaboration),
 Phys. Rev. Lett. {\bf 112}, 241101 (2014).

\bibitem{Note1}
Although the recent joint analysis \cite {Ade:2015tva} does not support this
  result, the discussion and the main results of this paper are not affected.

\bibitem{Kallosh:2013tua}
R.~Kallosh, A.~Linde, and D.~Roest,
 Phys. Rev. Lett. {\bf 112}, 011303 (2014).

\bibitem{Boubekeur:2005zm}
L.~Boubekeur and D.~Lyth,
 JCAP {\bf 0507}, 010 (2005).

\bibitem{Linde:1993cn}
A.~D. Linde,
 Phys. Rev. D {\bf 49}, 748 (1994).

\bibitem{Dvali:1994ms}
G.~Dvali, Q.~Shafi, and R.~K. Schaefer,
 Phys. Rev. Lett. {\bf 73}, 1886 (1994).

\bibitem{Copeland:1994vg}
E.~J. Copeland, A.~R. Liddle, D.~H. Lyth, E.~D. Stewart, and D.~Wands,
 Phys. Rev. D {\bf 49}, 6410 (1994).

\bibitem{Burgess:2013sla}
C.~Burgess, M.~Cicoli, and F.~Quevedo,
 JCAP {\bf 1311}, 003 (2013).

\bibitem{Gong:2014cqa}
Q.~Gao and Y.~Gong,
 Phys. Lett. B {\bf 734}, 41 (2014).

\bibitem{Lyth:1996im}
D.~H. Lyth,
 Phys. Rev. Lett. {\bf 78}, 1861 (1997).

\bibitem{BenDayan:2009kv}
I.~Ben-Dayan and R.~Brustein,
 JCAP {\bf 1009}, 007 (2010).

\bibitem{Lizarraga:2014eaa}
J.~Lizarraga {\em et~al.},
 Phys. Rev. Lett. {\bf 112}, 171301 (2014).

\bibitem{Harigaya:2014qza}
K.~Harigaya and T.~T. Yanagida,
 Phys. Lett. B {\bf 734}, 13 (2014).

\bibitem{Contaldi:2014zua}
C.~R. Contaldi, M.~Peloso, and L.~Sorbo,
 JCAP {\bf 1407}, 014 (2014).

\bibitem{Collins:2014yua}
H.~Collins, R.~Holman, and T.~Vardanyan,
 {Do Mixed States save Effective Field Theory from BICEP?},
 \href{http://arxiv.org/abs/1403.4592}{{\ttfamily arXiv:1403.4592}},
  2014.

\bibitem{Byrnes:2014xua}
C.~T. Byrnes, M.~Cortês, and A.~R. Liddle,
 Phys. Rev. D {\bf 90}, 023523 (2014).

\bibitem{Anchordoqui:2014uua}
L.~A. Anchordoqui, V.~Barger, H.~Goldberg, X.~Huang, and D.~Marfatia,
 Phys. Lett. B {\bf 734}, 134 (2014).

\bibitem{Harigaya:2014sua}
K.~Harigaya, M.~Ibe, K.~Schmitz, and T.~T. Yanagida,
 Phys. Lett. B {\bf 733}, 283 (2014).

\bibitem{Nakayama:2014koa}
K.~Nakayama and F.~Takahashi,
 Phys. Lett. B {\bf 734}, 96 (2014).

\bibitem{Zhao:2014rna}
W.~Zhao, C.~Cheng, and Q.-G. Huang,
 {Hint of relic gravitational waves in the Planck and WMAP data},
 \href{http://arxiv.org/abs/1403.3919}{{\ttfamily arXiv:1403.3919}}, 2014.

\bibitem{Cook:2014dga}
J.~L. Cook, L.~M. Krauss, A.~J. Long, and S.~Sabharwal,
 Phys. Rev. D {\bf 89}, 103525 (2014).

\bibitem{Miranda:2014wga}
V.~Miranda, W.~Hu, and P.~Adshead,
 Phys. Rev. D {\bf 89}, 101302 (2014).

\bibitem{Masina:2014yga}
I.~Masina,
 Phys. Rev. D {\bf 89}, 123505 (2014).

\bibitem{Hamada:2014iga}
Y.~Hamada, H.~Kawai, K.-y. Oda, and S.~C. Park,
 Phys. Rev. Lett. {\bf 112}, 241301 (2014).

\bibitem{Hertzberg:2014aha}
M.~P. Hertzberg,
 {Inflation, Symmetry, and B-Modes},
 \href{http://arxiv.org/abs/1403.5253}{{\ttfamily arXiv:1403.5253}},
  2014.

\bibitem{Dent:2014rga}
J.~B. Dent, L.~M. Krauss, and H.~Mathur,
 Phys. Lett. B {\bf 736}, 305 (2014).

\bibitem{Joergensen:2014rya}
J.~Joergensen, F.~Sannino, and O.~Svendsen,
 Phys. Rev. D {\bf 90}, 043509 (2014).

\bibitem{Freese:2014nla}
K.~Freese and W.~H. Kinney,
 {Natural Inflation: Consistency with Cosmic Microwave Background Observations
  of Planck and BICEP2},
 \href{http://arxiv.org/abs/1403.5277}{{\ttfamily arXiv:1403.5277}}, 2014.

\bibitem{Czerny:2014wua}
M.~Czerny, T.~Kobayashi, and F.~Takahashi,
 Phys. Lett. B {\bf 735}, 176 (2014).

\bibitem{Ferrara:2014ima}
S.~Ferrara, A.~Kehagias, and A.~Riotto,
 Fortsch. Phys. {\bf 62}, 573 (2014).

\bibitem{Zhu:2014wda}
T.~Zhu and A.~Wang,
 Phys. Rev. D {\bf 90}, 027304 (2014).

\bibitem{Okada:2014lxa}
N.~Okada, V.~N. Şenoğuz, and Q.~Shafi,
 {Simple Inflationary Models in Light of BICEP2: an Update},
 \href{http://arxiv.org/abs/1403.6403}{{\ttfamily arXiv:1403.6403}},
  2014.

\bibitem{Ellis:2014rxa}
J.~Ellis, M.~A. García, D.~V. Nanopoulos, and K.~A. Olive,
 JCAP {\bf 1405}, 037 (2014).

\bibitem{Kawai:2014doa}
S.~Kawai and N.~Okada,
 Phys. Lett. B {\bf 735}, 186 (2014).

\bibitem{Antusch:2014cpa}
S.~Antusch and D.~Nolde,
 JCAP {\bf 1405}, 035 (2014).

\bibitem{Freivogel:2014hca}
B.~Freivogel, M.~Kleban, M.~R. Martinez, and L.~Susskind,
 {Observational Consequences of a Landscape: Epilogue},
 \href{http://arxiv.org/abs/1404.2274}{{\ttfamily arXiv:1404.2274}}, 2014.

\bibitem{Bousso:2014jca}
R.~Bousso, D.~Harlow, and L.~Senatore,
 JCAP {\bf 1412}, 019 (2014).

\bibitem{Kaloper:2014zba}
N.~Kaloper and A.~Lawrence,
 Phys. Rev. D {\bf 90}, 023506 (2014).

\bibitem{Choudhury:2013iaa}
S.~Choudhury and A.~Mazumdar,
 Nucl. Phys. B {\bf 882}, 386 (2014).

\bibitem{Choudhury:2014wsa}
S.~Choudhury and A.~Mazumdar,
 {Sub-Planckian inflation \& large tensor to scalar ratio with $r\geq 0.1$},
 \href{http://arxiv.org/abs/1404.3398}{{\ttfamily arXiv:1404.3398}},
  2014.

\bibitem{Choudhury:2014kma}
S.~Choudhury and A.~Mazumdar,
 {Reconstructing inflationary potential from BICEP2 and running of tensor
  modes},
 \href{http://arxiv.org/abs/1403.5549}{{\ttfamily arXiv:1403.5549}},
  2014.

\bibitem{Choudhury:2014uxa}
S.~Choudhury,
 JHEP {\bf 04}, 105 (2014).

\bibitem{Choi:2014aca}
K.-Y. Choi and B.~Kyae,
 Phys. Lett. B {\bf 735}, 391 (2014).

\bibitem{Murayama:2014saa}
H.~Murayama, K.~Nakayama, F.~Takahashi, and T.~T. Yanagida,
 Phys. Lett. B {\bf 738}, 196 (2014).

\bibitem{McDonald:2014oza}
J.~McDonald,
 JCAP {\bf 1409}, 027 (2014).

\bibitem{Gao:2014fha}
X.~Gao, T.~Li, and P.~Shukla,
 Phys. Lett. B {\bf 738}, 412 (2014).

\bibitem{Li:2014owa}
T.~Li, Z.~Li, and D.~V. Nanopoulos,
 {Chaotic Inflation in No-Scale Supergravity with String Inspired Moduli
  Stabilization},
 \href{http://arxiv.org/abs/1405.0197}{{\ttfamily arXiv:1405.0197}},
  2014.

\bibitem{Li:2014xna}
T.~Li, Z.~Li, and D.~V. Nanopoulos,
 JHEP {\bf 1407}, 052 (2014).

\bibitem{Ashoorioon:2014nta}
A.~Ashoorioon, K.~Dimopoulos, M.~Sheikh-Jabbari, and G.~Shiu,
 Phys. Lett. B {\bf 737}, 98 (2014).

\bibitem{Ashoorioon:2013eia}
A.~Ashoorioon, K.~Dimopoulos, M.~Sheikh-Jabbari, and G.~Shiu,
 JCAP {\bf 1402}, 025 (2014).

\bibitem{Ashoorioon:2009wa}
A.~Ashoorioon, H.~Firouzjahi, and M.~Sheikh-Jabbari,
 JCAP {\bf 0906}, 018 (2009).

\bibitem{Ashoorioon:2011ki}
A.~Ashoorioon and M.~Sheikh-Jabbari,
 JCAP {\bf 1106}, 014 (2011).

\bibitem{Sloth:2014sga}
M.~S. Sloth, Phys. Rev. D {\bf 90}, 063511 (2014).
  2014.

\bibitem{Kobayashi:2014rla}
T.~Kobayashi and O.~Seto,
 {Beginning of Universe through large field hybrid inflation},
 \href{http://arxiv.org/abs/1404.3102}{{\ttfamily arXiv:1404.3102}},
  2014.

\bibitem{Kobayashi:2014ooa}
T.~Kobayashi, O.~Seto, and Y.~Yamaguchi,
 Prog. Theor. Exp. Phys. {\bf 2014}, 103E01 (2014).

\bibitem{Kobayashi:2014jga}
T.~Kobayashi and O.~Seto,
 Phys. Rev. D {\bf 89}, 103524 (2014).

\bibitem{Bastero-Gil:2014oga}
M.~Bastero-Gil, A.~Berera, R.~O. Ramos, and J.~G. Rosa,
 JCAP {\bf 1410}, 053 (2014).

\bibitem{DiBari:2014oja}
P.~Di~Bari, S.~F. King, C.~Luhn, A.~Merle, and A.~Schmidt-May,
 JCAP {\bf 1408}, 040 (2014).

\bibitem{Ho:2014xza}
C.~M. Ho and S.~D.~H. Hsu,
 JHEP {\bf 1407}, 060 (2014).

\bibitem{Hotchkiss:2011gz}
S.~Hotchkiss, A.~Mazumdar, and S.~Nadathur,
 JCAP {\bf 1202}, 008 (2012).

\bibitem{Dvali:2014ssa}
G.~Dvali and C.~Gomez,
 {BICEP2 in Corpuscular Description of Inflation},
 \href{http://arxiv.org/abs/1403.6850}{{\ttfamily arXiv:1403.6850}}, 2014.

\bibitem{Palti:2014kza}
E.~Palti and T.~Weigand,
 JHEP {\bf 1404}, 155 (2014).

\bibitem{Ma:2014vua}
Y.-Z. Ma and Y.~Wang,
 JCAP {\bf 1409}, 041 (2014).

\bibitem{Hamaguchi:2014mza}
K.~Hamaguchi, T.~Moroi, and T.~Terada,
 Phys. Lett. B {\bf 733}, 305 (2014).

\bibitem{Creminelli:2014oaa}
P.~Creminelli, D.~López~Nacir, M.~Simonović, G.~Trevisan, and M.~Zaldarriaga,
 Phys. Rev. Lett. {\bf 112}, 241303 (2014).

\bibitem{Kamionkowski:2014faa}
M.~Kamionkowski, L.~Dai, and D.~Jeong,
 Phys. Rev. D {\bf 89}, 107302 (2014).

\bibitem{Carrillo-Gonzalez:2014tia}
M.~Carrillo-González, G.~Germán-Velarde, A.~Herrera-Aguilar, J.~C. Hidalgo,
  and R.~Sussman,
 Phys. Lett. B {\bf 734}, 345 (2014).

\bibitem{delCampo:2014toa}
S.~del Campo,
 {Intermediate inflation under the scrutiny of recent data},
 \href{http://arxiv.org/abs/1404.1649}{{\ttfamily arXiv:1404.1649}}, 2014.

\bibitem{Marchesano:2014mla}
F.~Marchesano, G.~Shiu, and A.~M. Uranga,
 JHEP {\bf 1409}, 184 (2014).

\bibitem{Choi:2014dva}
K.-Y. Choi and B.~Kyae,
 Phys. Rev. D {\bf 90}, 023536 (2014).

\bibitem{Oda:2014rpa}
I.~Oda and T.~Tomoyose,
 Adv. Stud. Theor. Phys. {\bf 8}, 551 (2014).

\bibitem{Yonekura:2014oja}
K.~Yonekura,
 JCAP {\bf 1410}, 054 (2014).

\bibitem{Higaki:2014sja}
T.~Higaki, T.~Kobayashi, O.~Seto, and Y.~Yamaguchi,
 JCAP {\bf 1410}, 025 (2014).

\bibitem{Bamba:2014jia}
K.~Bamba, R.~Myrzakulov, S.~Odintsov, and L.~Sebastiani,
 Phys. Rev. D {\bf 90}, 043505 (2014).

\bibitem{Hebecker:2014eua}
A.~Hebecker, S.~C. Kraus, and L.~T. Witkowski,
 Phys. Lett. B {\bf 737}, 16 (2014).

\bibitem{Long:2014dta}
C.~Long, L.~McAllister, and P.~McGuirk,
 Phys. Rev. D {\bf 90}, 023501 (2014).

\bibitem{Ben-Dayan:2014zsa}
I.~Ben-Dayan, F.~G. Pedro, and A.~Westphal,
 Phys. Rev. Lett. {\bf 113}, 261301 (2014).

\bibitem{Freedman:1976xh}
D.~Z. Freedman, P.~van Nieuwenhuizen, and S.~Ferrara,
 Phys. Rev. D {\bf 13}, 3214 (1976).

\bibitem{Deser:1976eh}
S.~Deser and B.~Zumino,
 Phys. Lett. B {\bf 62}, 335 (1976).

\bibitem{Antusch:2009ty}
S.~Antusch, M.~Bastero-Gil, K.~Dutta, S.~F. King, and P.~M. Kostka,
 Phys. Lett. B {\bf 679}, 428 (2009).

\bibitem{Antusch:2011ei}
S.~Antusch, K.~Dutta, J.~Erdmenger, and S.~Halter,
 JHEP {\bf 1104}, 065 (2011).

\bibitem{Stewart:1994ts}
E.~D. Stewart,
 Phys. Rev. D {\bf 51}, 6847 (1995).

\bibitem{adlinde90}
A.~Linde,
 {\em Particle Physics and Inflationary Cosmology} (Harwood Academic, Chur, Switzerland, 1990).

\bibitem{Antusch:2008pn}
S.~Antusch, M.~Bastero-Gil, K.~Dutta, S.~F. King, and P.~M. Kostka,
 JCAP {\bf 0901}, 040 (2009).

\bibitem{Yamaguchi:2011kg}
M.~Yamaguchi,
 Class. Quant. Grav. {\bf 28}, 103001 (2011).

\bibitem{Martin:2014vha}
J.~Martin, C.~Ringeval, and V.~Vennin,
 Phys. Dark Univ. {\bf 5-6}, 75-235 (2014).

\bibitem{Lyth:1998xn}
D.~H. Lyth and A.~Riotto,
 Phys. Rept. {\bf 314}, 1 (1999).

\bibitem{Goncharov:1984qm}
A.~S. Goncharov, A.~D. Linde, and M.~I. Vysotsky,
 Phys. Lett. B {\bf 147}, 279 (1984).

\bibitem{Cremmer:1983bf}
E.~Cremmer, S.~Ferrara, C.~Kounnas, and D.~V. Nanopoulos,
 Phys. Lett. B {\bf 133}, 61 (1983).

\bibitem{Ellis:1983sf}
J.~R. Ellis, A.~Lahanas, D.~V. Nanopoulos, and K.~Tamvakis,
 Phys. Lett. B {\bf 134}, 429 (1984).

\bibitem{Ellis:1983ei}
J.~R. Ellis, C.~Kounnas, and D.~V. Nanopoulos,
 Nucl. Phys. B {\bf 241}, 406 (1984).

\bibitem{Ellis:1984bm}
J.~R. Ellis, C.~Kounnas, and D.~V. Nanopoulos,
 Nucl. Phys. B {\bf 247}, 373 (1984).

\bibitem{Lahanas:1986uc}
A.~Lahanas and D.~V. Nanopoulos,
 Phys. Rept. {\bf 145}, 1 (1987).

\bibitem{Ellis:1984bf}
J.~R. Ellis, K.~Enqvist, D.~V. Nanopoulos, K.~A. Olive, and M.~Srednicki,
 Phys. Lett. B {\bf 152}, 175 (1985); {\bf 156}, 452(E) (1985).

\bibitem{Enqvist:1985yc}
K.~Enqvist, D.~V. Nanopoulos, and M.~Quiros,
 Phys. Lett. B {\bf 159}, 249 (1985).

\bibitem{Ellis:2013xoa}
J.~Ellis, D.~V. Nanopoulos, and K.~A. Olive,
 Phys. Rev. Lett. {\bf 111}, 111301 (2013); {\bf 111}, 129902 (2013).

\bibitem{Ellis:2013nxa}
J.~Ellis, D.~V. Nanopoulos, and K.~A. Olive,
 JCAP {\bf 1310}, 009 (2013).

\bibitem{Li:2013moa}
T.~Li, Z.~Li, and D.~V. Nanopoulos,
 JCAP {\bf 1404}, 018 (2014).

\bibitem{Ellis:2013nka}
J.~Ellis, D.~V. Nanopoulos, and K.~A. Olive,
 Phys. Rev. D {\bf 89}, 043502 (2014).

\bibitem{Kawasaki:2000yn}
M.~Kawasaki, M.~Yamaguchi, and T.~Yanagida,
 Phys. Rev. Lett. {\bf 85}, 3572 (2000).

\bibitem{Yamaguchi:2000vm}
M.~Yamaguchi and J.~Yokoyama,
 Phys. Rev. D {\bf 63}, 043506 (2001).

\bibitem{Yamaguchi:2001pw}
M.~Yamaguchi,
 Phys. Rev. D {\bf 64}, 063502 (2001).

\bibitem{Kawasaki:2001as}
M.~Kawasaki and M.~Yamaguchi,
 Phys. Rev. D {\bf 65}, 103518 (2002).

\bibitem{Kallosh:2010ug}
R.~Kallosh and A.~Linde,
 JCAP {\bf 1011}, 011 (2010).

\bibitem{Kallosh:2010xz}
R.~Kallosh, A.~Linde, and T.~Rube,
 Phys. Rev. D {\bf 83}, 043507 (2011).

\bibitem{Nakayama:2013jka}
K.~Nakayama, F.~Takahashi, and T.~T. Yanagida,
 Phys. Lett. B {\bf 725}, 111 (2013).

\bibitem{Nakayama:2013txa}
K.~Nakayama, F.~Takahashi, and T.~T. Yanagida,
 JCAP {\bf 1308}, 038 (2013).

\bibitem{Takahashi:2013cxa}
F.~Takahashi,
 Phys. Lett. B {\bf 727}, 21 (2013).

\bibitem{Li:2013nfa}
T.~Li, Z.~Li, and D.~V. Nanopoulos,
 JCAP {\bf 1402}, 028 (2014).

\bibitem{Stewart:1993bc}
E.~D. Stewart and D.~H. Lyth,
 Phys. Lett. B {\bf 302}, 171 (1993).

\bibitem{Gao:2014yra}
Q.~Gao, Y.~Gong, T.~Li, and Y.~Tian,
 Sci. China Phys. Mech. Astron. {\bf 57}, 1442 (2014).

\bibitem{Hebecker:2013zda}
A.~Hebecker, S.~C. Kraus, and A.~Westphal,
 Phys. Rev. D {\bf 88}, 123506 (2013).

\bibitem{German:2014qza}
G.~German,
 {On the Lyth bound and single-field slow-roll inflation},
 \href{http://arxiv.org/abs/1405.3246}{{\ttfamily arXiv:1405.3246}}, 2014.

\bibitem{Efstathiou:2005tq}
G.~Efstathiou and K.~J. Mack,
 JCAP {\bf 0505}, 008 (2005).

\bibitem{Gong:2001he}
J.-O. Gong and E.~D. Stewart,
 Phys. Lett. B {\bf 510}, 1 (2001).

\bibitem{Zhu:2014aea}
T.~Zhu, A.~Wang, G.~Cleaver, K.~Kirsten, and Q.~Sheng,
 Phys. Rev. D {\bf 90}, 063503 (2014).

\bibitem{Mukhanov:1985rz}
V.~F. Mukhanov,
 JETP Lett. {\bf 41}, 493 (1985).

\bibitem{Sasaki:1986hm}
M.~Sasaki,
 Prog. Theor. Phys. {\bf 76}, 1036 (1986).

\bibitem{Ade:2015tva} P.~Ade {\em et~al.} (BICEP2
  Collaboration, Planck Collaboration),
 {A Joint Analysis of BICEP2/Keck Array and Planck Data},
 \href{http://arxiv.org/abs/1502.00612}{{\ttfamily arXiv:1502.00612}}, 2015.

\end{thebibliography}

\end{document}